\documentclass[reprint,superscriptaddress,amsmath,amssymb,aps]{revtex4-2}
\usepackage{natbib}
\usepackage[colorlinks,linkcolor = blue,citecolor=blue,urlcolor=blue,bookmarks=false,hypertexnames=true]{hyperref}
\usepackage{float}
\usepackage{comment}
\usepackage{mathtools}
\usepackage{graphicx}
\usepackage[export]{adjustbox}
\usepackage{dcolumn}
\usepackage{bm}
\usepackage{sidecap}
\usepackage{chemformula}
\usepackage{textcomp}
\usepackage[justification=justified,singlelinecheck=false]{subcaption}
\usepackage{gensymb}
\usepackage{subfiles}

\begin{document}

\title{Magnon-polaron mediated spin Seebeck effect in altermagnets}

\author{Ilia Moghayer}
\affiliation{%
Kavli Institute of Nanoscience, Delft University of Technology, Lorentzweg 1, 2628 CJ, Delft, The
Netherlands
}

\affiliation{Institute for Theoretical Physics, ETH Zurich, CH-8093 Zurich, Switzerland}%

\author{Ritesh Das}
\email{R.D.Das@tudelft.nl}
\affiliation{%
Kavli Institute of Nanoscience, Delft University of Technology, Lorentzweg 1, 2628 CJ, Delft, The
Netherlands
}%

\author{Yaroslav M. Blanter }
\email{Y.M.Blanter@tudelft.nl}
\affiliation{%
Kavli Institute of Nanoscience, Delft University of Technology, Lorentzweg 1, 2628 CJ, Delft, The
Netherlands
}%

\date{\today}

\begin{abstract}
Altermagnets, distinguished by compensated antiparallel spins yet nondegenerate magnon spectra, bridge the gap between ferromagnets and antiferromagnets. Although several probes such as anisotropic transport and spectroscopic measurements have been proposed to identify altermagnetic order, experimentally accessible transport signatures remain highly desirable. Here, we show that in altermagnets subject to an external magnetic field, the spin Seebeck effect exhibits pronounced directional anisotropy. Specifically, the spin Seebeck coefficient differs significantly along in-plane directions, and this anisotropy increases with field strength. Magnetoelastic interactions further produce resonant peaks whose positions with respect to an applied magnetic field reflect the intrinsic magnon band anisotropy, and provide localized features that enhance the distinguishability of the altermagnetic response. The peaks provide a robust experimental signature of altermagnetic order. Our findings serve as signatures of altermagnetic order while laying the groundwork for their application in spintronic devices.

\end{abstract}

\maketitle

Interest in magnetism has grown substantially due to the rich variety of existing magnetic phases, each distinguished by the way microscopic spins organize and interact \cite{stohr2006magnetism, stamps20142014,stancil2009spin, Rezende2020}. Ferromagnets are characterized by parallel alignment of spins, resulting in a finite net magnetization \cite{stancil2009spin, Rezende2020, Holstein1940, Landau1935,gilbert2004phenomenological}. Their long-range order underpins both fundamental studies and technological applications, most notably in magnetic storage and spintronic devices. Antiferromagnets, in contrast, host antiparallel spin alignment with a vanishing macroscopic magnetization \cite{Rezende2019, alaei2025origin, yang2024exchange,gray2019spin, xu2022observation, dzian2025antiferromagnetic, rezende2016theory}. This absence of net moment provides robustness against stray fields and ultrafast spin dynamics, but simultaneously complicates their detection as conventional magnetometry is largely insensitive to compensated spin order. Altermagnets, recently identified as a distinct class of collinear magnets, provide a new paradigm in this landscape \cite{song2025altermagnets, vsmejkal2022beyond, vsmejkal2022emerging,mcclarty2024landau,vsmejkal2023chiral}. They combine compensated antiparallel spin order with a symmetry-protected lifting of magnon (and electronic) band degeneracy \cite{krempasky2024altermagnetic, sodequist2024two, vsmejkal2023chiral}. As such, altermagnets occupy an intermediate position between ferromagnets and antiferromagnets as they host no net magnetization, yet exhibit spin-polarized transport channels. 

The significance of altermagnets is twofold. From a practical perspective, they promise spintronic functionality akin to ferromagnets \cite{vzutic2004spintronics, yi2020concepts, puebla2020spintronic} without producing stray fields, thus enabling compact device architectures with reduced cross-talk. From a fundamental standpoint, they extend the classification of magnetic order and provide a natural platform for studying unconventional quasiparticles and anisotropic spin transport. However, their experimental identification remains challenging. The absence of net magnetization prevents most conventional detection schemes. At the same time, direct observation of their symmetry-induced band splitting often requires momentum-resolved probes such as angle-resolved photoemission spectroscopy \cite{reimers2024direct} or inelastic neutron scattering \cite{neutronscattering1,mcclarty2025observing, faure2025altermagnetism}. This motivates the development of alternative, transport-based probes. Anisotropies in the spin transport quantities can potentially offer unambiguous signature that differentiates altermagnets from conventional ferro- and antiferromagnets. One such quantity is the spin Seebeck effect (SSE) - the conversion of a temperature gradient into a spin current - which has been established as a central platform for probing magnon-mediated spin transport in ferromagnets \cite{cornelissen2017nonlocal}. In collinear antiferromagnets, magnon bands associated with opposite spin angular momentum are typically degenerate across the Brillouin zone. As a consequence, the SSE vanishes at zero magnetic field unless this degeneracy is lifted by spin canting, external fields, or Dzyaloshinskii–Moriya interactions. However, magnon bands in altermagnets remain non-degenerate even in the absence of spin–orbit coupling or external fields, enabling a finite, anisotropic SSE at zero field, and providing a unique signature to distinguish them from antiferromagnets.

An additional layer of complexity arises from magnetoelastic interactions, i.e., the coupling between magnons and lattice vibrations (phonons). Under specific field configurations, magnon and phonon modes hybridize to form magnon–polarons, producing pronounced enhancements of the SSE \cite{cornelissen2017nonlocal,kikkawa2016magnon,flebus2017magnon}. While this effect has been extensively investigated in ferromagnets and to a lesser extent in antiferromagnets \cite{liu2022magnon}, its role in altermagnets has not, to the best of our knowledge, been explored. In this work, we show that the anisotropic SSE effect in altermagnets, enhanced by the magnetoelastic interaction, can be used to distinguish altermagnets from antiferromagnets.

In experimental scenarios, the measured response can contain multiple spin-dependent transport mechanisms. These contributions typically vary smoothly with magnetic field and can obscure the interpretation of broad spin Seebeck anisotropies. 
In contrast, magnon–phonon hybridization gives rise to localized peaks in the SSE as a function of magnetic field, occurring at the resonant fields where the magnon and phonon dispersions intersect. Because the positions of these resonances can be predicted from the magnon and phonon spectra, such peaks provide a more robust diagnostic signature even in the presence of additional background contributions. Phonon-mediated SSE can be competitive with neutron spectroscopy as a probe of altermagnetic order. For example, in the candidate altermagnet MnF$_2$, whose predicted splitting has so far eluded inelastic neutron scattering \cite{morano2025absence}, the theoretical magnon–polaron resonance fields of the two branches remain resolvable below the neutron sensitivity bound, suggesting the viability of our proposed method (a deeper discussion is provided in section IV of the supplementary information).

In this article, we investigate the spin Seebeck effect in a minimal toy model of an altermagnet as a function of external magnetic field in the presence of phonons. The magnetic field is applied along the same direction in which the SSE is measured. We demonstrate that the SSE response in altermagnets exhibits strong anisotropy with respect to the direction of the applied field even in the absence of phonons. A stronger signature of altermagnetic character is seen in the SSE in the presence of phonons. A peak is expected in the SSE at the field where the magnon and phonon modes hybridize \cite{kikkawa2016magnon,flebus2017magnon,cornelissen2017nonlocal}. In an antiferromagnet, a peak is seen in the magnon-polaron-mediated spin Seebeck coefficient at the same magnetic field along all directions. In contrast, in an altermagnet, the peaks in the SSE in different directions appear on applying different strengths of magnetic field, reflecting the intrinsic anisotropy of magnon dispersion when comparing measurements of the SSE along different directions. This can act as a direct probe to distinguish altermagnets from antiferromagnets. 

To capture the essential physics of an altermagnet, we consider a minimal lattice model shown in Fig. 1 (a) comprising two sublattices with antiparallel spins. Neighboring spins from opposite sublattices are coupled antiferromagnetically (exchange $J_{1}$) while next-nearest neighbors on the same sublattice have ferromagnetic couplings ($J_{2}$ along one crystal axis and $J'_{2}$ along the other). $J_{2} \neq J'_{2}$ enables altermagnetism while $J_{2} = J_{2}'$ corresponds to antiferromagnetic behavior. This monolayer structure is seen in Cr$_{2}$Te$_{2}$O \cite{Cui2023}.  
\begin{figure}[]
 \begin{center}
    \begin{subfigure}{0.23\textwidth}
    \includegraphics[width=\linewidth]{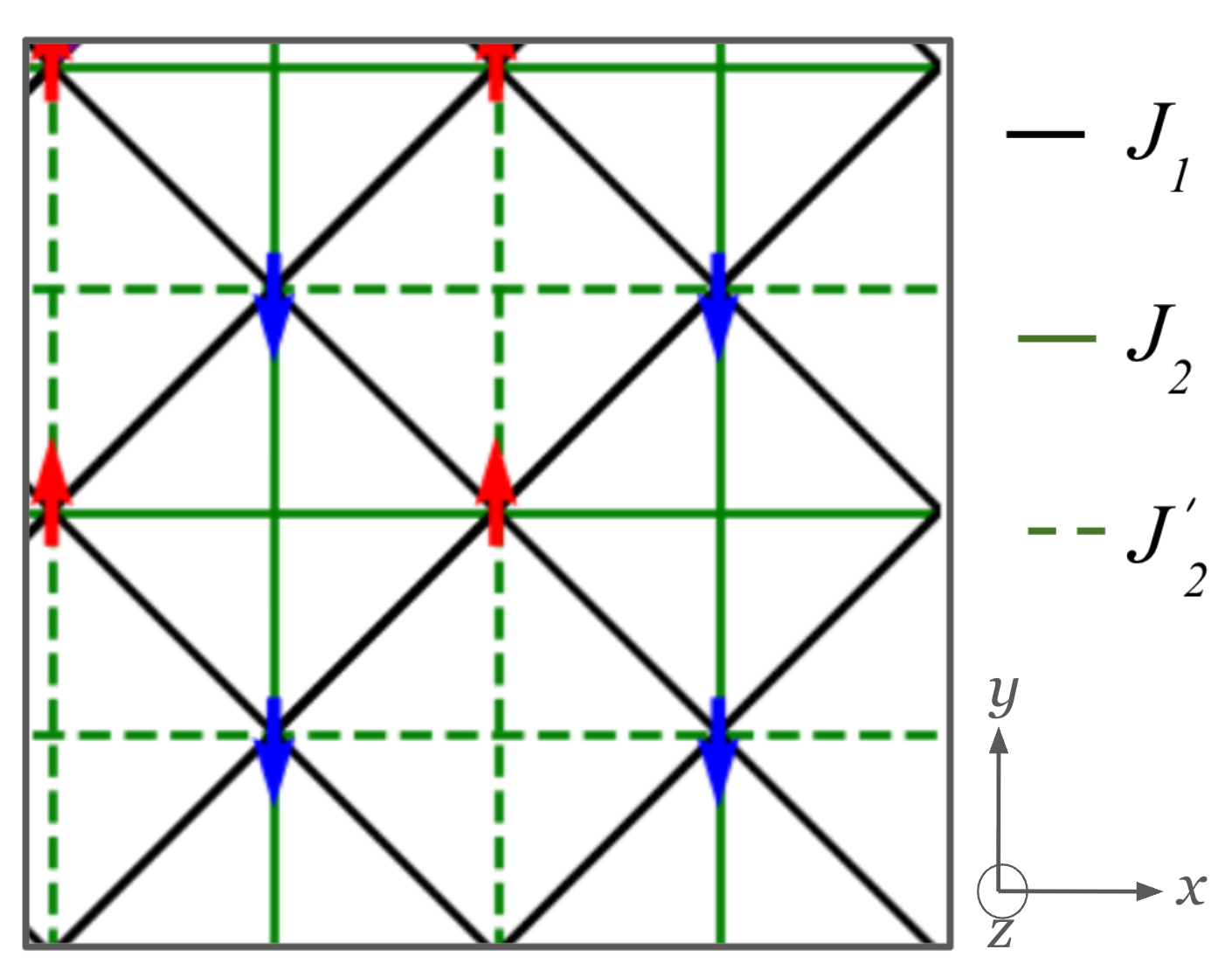}
    \caption{} 
  \end{subfigure}
  \begin{subfigure}{0.22\textwidth}
    \includegraphics[width=\linewidth]{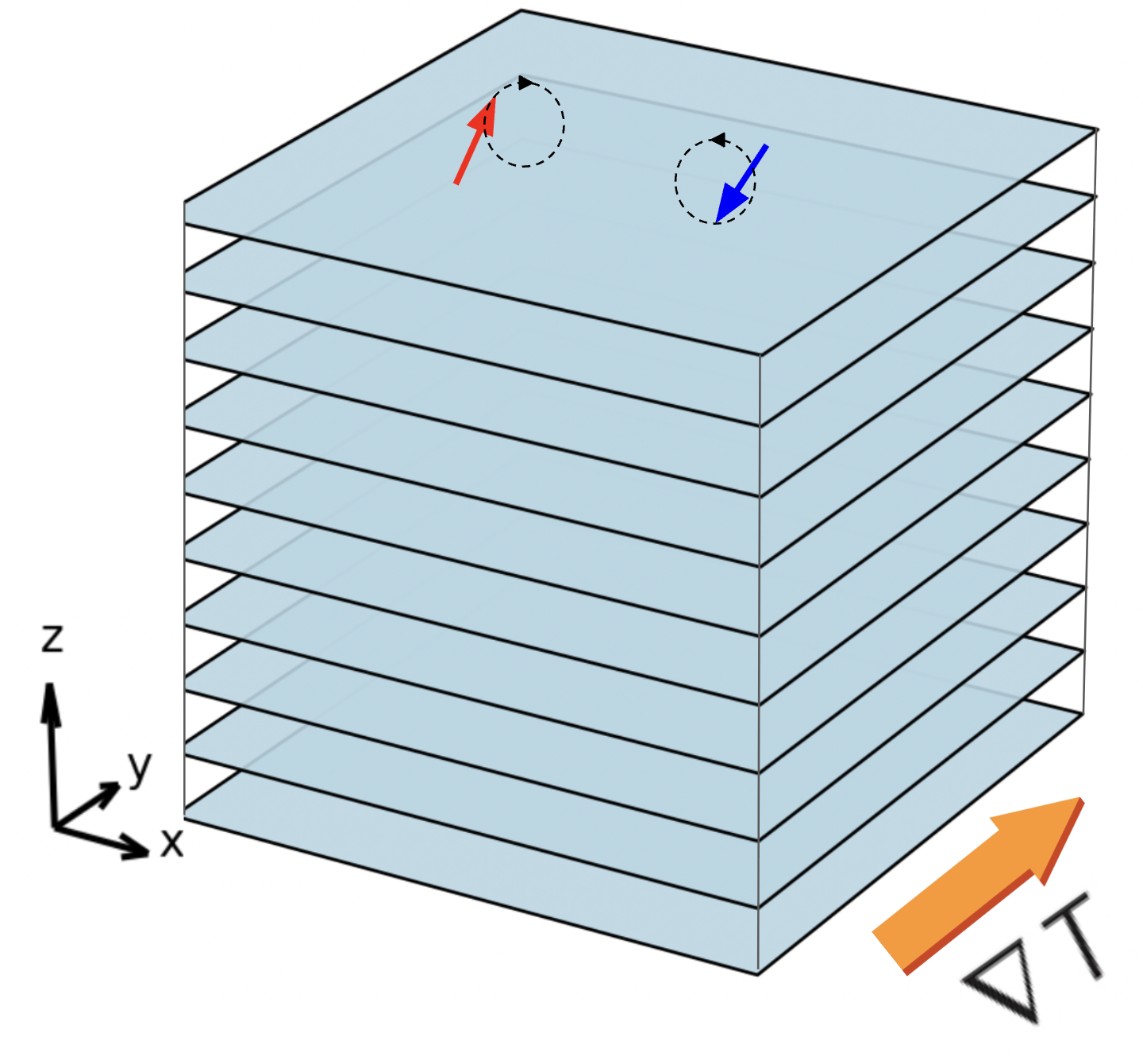}
    \caption{} 
  \end{subfigure}
  \end{center}
    \begin{center}
 \caption{\centerlast (a) Lattice structure of the altermagnetic toy model. The lattice has a checkerboard shape with sites alternating between up and down spins. The altermagnetic nature arises from the anisotropic next-nearest neighbor interaction ($J_{2} \neq J'_{2}$). (b) Schematic of spin Seebeck effect in the bulk altermagnetic system. The bulk consists of multiple altermagnetic monolayers coupled ferromagnetically to each other. A temperature gradient is applied along direction $\beta$ (shown along $y$ here) and the spin Seebeck response is measured along direction $\alpha$.}
    \end{center}
 \end{figure}

We model spin dynamics with the Hamiltonian\cite{Cui2023, vsmejkal2022beyond}:
\begin{align}
\begin{split}
\hat{H} = &  \hspace{0.1cm} J_{1}\sum_{\langle i,j \rangle} \mathbf{S}_{a,i} \cdot \mathbf{S}_{b,j} 
- H_{0}  \sum_{i} \Big( \mathbf{S}_{a,i}^{y} + \mathbf{S}_{b,i}^{y} \Big) 
\\
& + J'_{2} \Big( \sum_{\langle\langle i,j \rangle\rangle_{y}} \mathbf{S}_{a,i} \cdot \mathbf{S}_{a,j} + \sum_{\langle\langle i,j \rangle\rangle_{x}} \mathbf{S}_{b,i} \cdot \mathbf{S}_{b,j} \Big)\\
&+ J_{2} \Big( \sum_{\langle\langle i,j \rangle\rangle_{y}} \mathbf{S}_{b,i} \cdot \mathbf{S}_{b,j}  +  \sum_{\langle\langle i,j \rangle\rangle_{x}} \mathbf{S}_{a,i} \cdot \mathbf{S}_{a,j} \Big) \\
& +D \sum_{i} [(S^{y}_{a,i})^{2} + (S^{y}_{b,i})^{2}].
\end{split}
\end{align}

Here, $a$ and $b$ denote the spin-up and spin-down sublattices, respectively, $\langle i,j\rangle$ denotes summation over all lattice sites $i$ and nearest neighbors $j$, $\langle \langle i,j\rangle \rangle_{\alpha}$ denotes summation over the next-nearest neighbors along the direction $\alpha \in \{ x, y\}$,
$J_{1}$ is the inter-sublattice nearest neighbor antiferromagnetic exchange, $J_{2}$ and $J_{2}'$ are the intra-sublattice next-nearest neighbor exchanges, $D$ is the easy-axis anisotropy along $y$ , and $H_{0}$ is a Zeeman term for an external field along $y$.

Diagonalizing the Hamiltonian in Eq. (1) yields two magnon modes $\omega_{A}(\mathbf{k})$ and $\omega_{B}(\mathbf{k})$ whose general form is given by:

\begin{equation}
\resizebox{0.9\hsize}{!}{$
\omega_{A,B} = \frac{ \mp (\Lambda_2 - \Lambda_1) + \sqrt{ (\Lambda_2 - \Lambda_1)^2 + 4(\Lambda_1 \Lambda_2 - X^2)} }{2},
$}
\end{equation}

where $\Lambda_{1,2},$ and $X$ are combinations of the Zeeman field, exchange parameters and anisotropy given in Equations (4) and (7) of the supplementary information.
In the altermagnetic phase ($J_{2} \ne J'_{2}$), one finds $\Lambda_{1} \ne \Lambda_{2}$, yielding $\omega_{A}(\mathbf{k}) \ne \omega_{B}(\mathbf{k})$ for every wavevector $\mathbf{k}$. In contrast, setting $J_{2} = J'_{2}$ results in $\Lambda_{1} = \Lambda_{2}$ in the absence of an external field, recovering two degenerate magnon modes $\omega_{A}(\mathbf{k})$ at $H_{0} = 0$. In this case, the model reduces to a square-lattice antiferromagnet with degenerate magnon branches. The presence of nondegenerate, direction-dependent magnon dispersion in altermagnets is the key distinction that enables a finite SSE without applied field. 

\begin{figure}[]
 \begin{center}
    \begin{subfigure}{0.23\textwidth}
    \includegraphics[width=\linewidth]{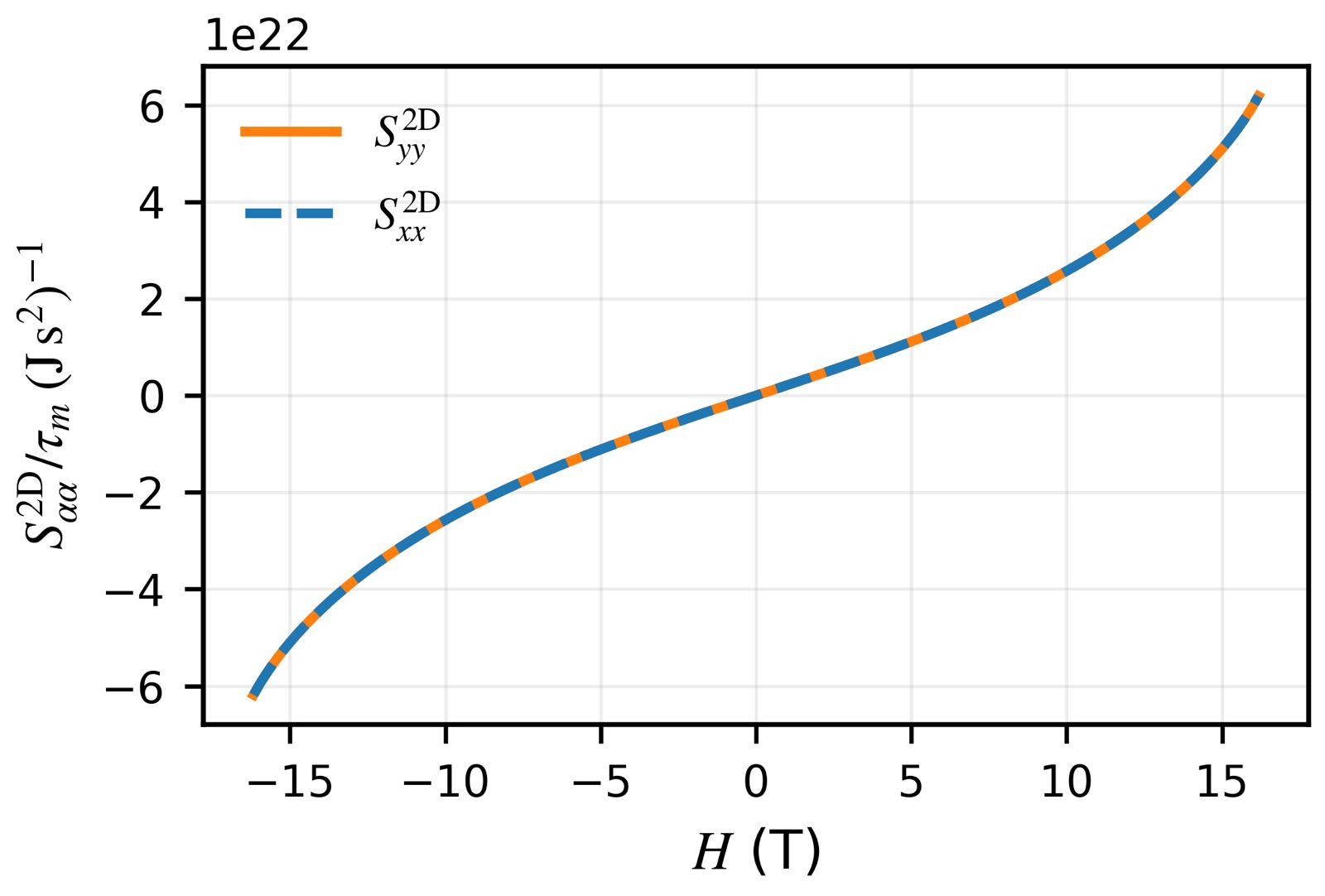}
    \caption{} 
  \end{subfigure}
  \begin{subfigure}{0.23\textwidth}
    \includegraphics[width=\linewidth]{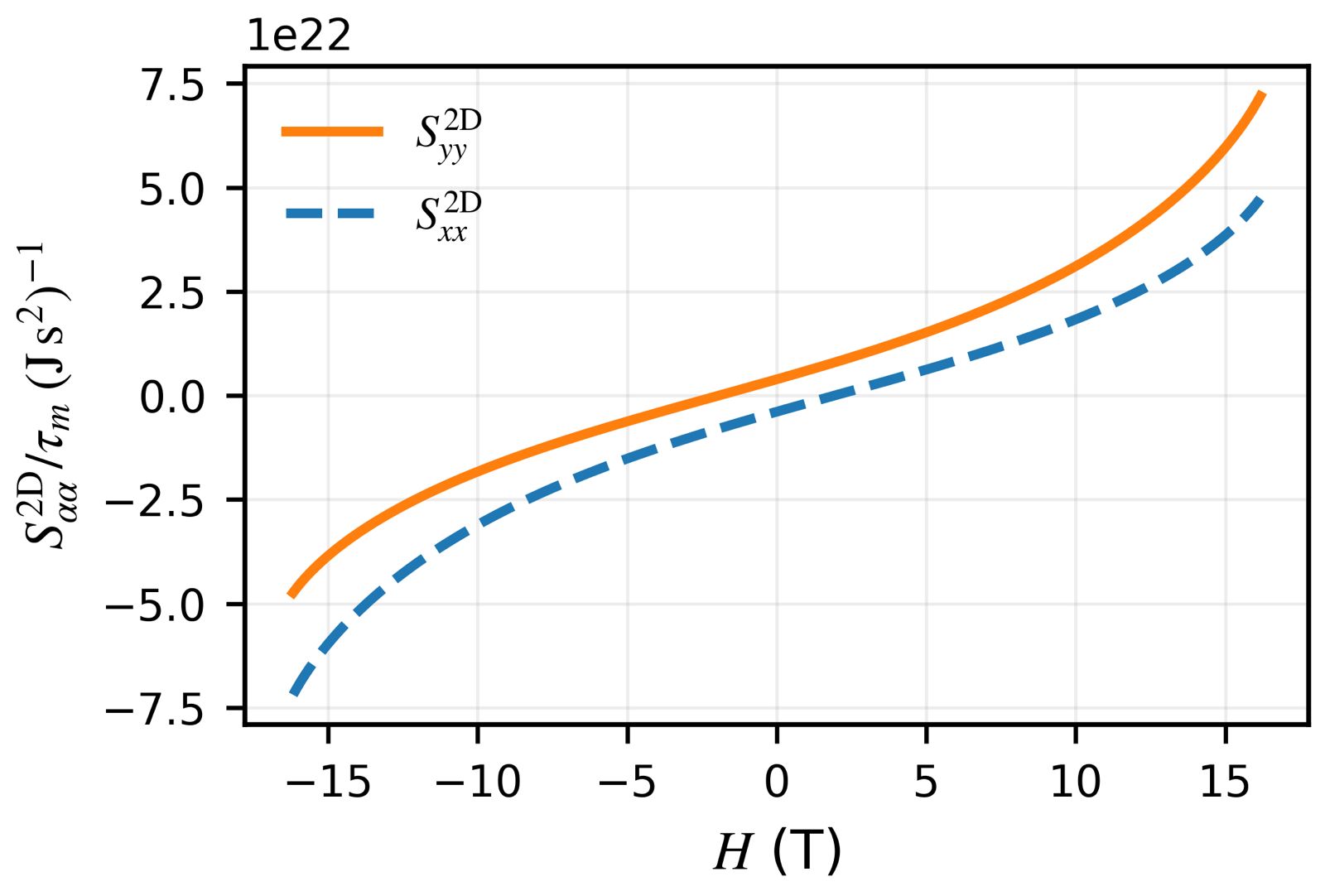}
    \caption{} 
  \end{subfigure}
  \end{center}
\caption{\centerlast Spin Seebeck coefficient $S^{2D}_{xx}$ and $S^{2D}_{yy}$ as a function of applied magnetic field without phonon contributions for (a) antiferromagnetic and (b) altermagnetic configurations in the 2D limit.}
 \end{figure}

The directional anisotropy in the spin Seebeck effect becomes even more pronounced when phonons are included in the system. Magnons and phonons can hybridize to form magnon-polarons. In the long-wavelength limit, the magnetoelastic energy per unit cell $n$ can be expressed in terms of small deviations of the Néel vector $\vec{L}^{(n)} = \mathbf{S}^{(n)}_{A} -\mathbf{S}^{(n)}_{B}$, where $\displaystyle \mathbf{S}^{(n)}_{A} = \sum_{i \in n} \mathbf{S}_{a,i} $ and $\displaystyle \mathbf{S}^{(n)}_{B} = \sum_{i \in n} \mathbf{S}_{b,i}$, and strain $u^{(n)}$ from equilibrium \cite{turov2001simmetriya}:
\begin{equation}
\begin{aligned}
E_{\text{me}} = \sum_{i} \sum_{\alpha\beta\gamma\delta} [
B_{\alpha\beta\gamma\delta} \, \vec{L}_\alpha \vec{L}_\beta \, u_{\gamma\delta}
+ B'_{\alpha\beta\gamma\delta} \, \frac{\partial \vec{L}}{\partial r_\alpha} \cdot \frac{\partial \vec{L}}{\partial r_\beta} \, u_{\gamma\delta}
]
\end{aligned}
\end{equation}
for $\alpha, \beta, \gamma, \delta \in \{x,y,z\}$, where $B_{\alpha \beta \gamma \delta}$ and $B'_{\alpha \beta \gamma \delta}$ are magnetoelastic coupling coefficients determined by crystal symmetry, $u_{\alpha\beta} = \frac{1}{2}(\partial_{\alpha} \mathbf{u}_{\beta} + \partial_{\beta} \mathbf{u}_{\alpha})$ is the strain tensor, $\mathbf{u}_{\alpha}$ being the displacement field along $\alpha$ (See supplementary information for details of the phonon modes). Here, we have only considered the linear resonant coupling between magnons and phonons. 

Next, we calculate the spin Seebeck coefficient $\mathcal{S}_{\alpha\beta}$, defined as the spin-current response measured along $\alpha$ for a temperature gradient applied along $\beta$ (See Fig. 1 (b)). Within Boltzmann theory, the SSE with hybridized magnon-polaron modes is given by an expression of the form \cite{flebus2017magnon,kikkawa2016magnon, cornelissen2017nonlocal}:

\begin{equation}
\begin{aligned}
S_{\alpha\beta} = -\frac{\hbar}{(2\pi)^n k_B T} \int d^n \mathbf{k} \sum_{i} W_{i}(\mathbf{k}) \mathcal{S}_{i}(\mathbf{k}) \tau_{i}(\mathbf{k})\\
\times v_{i,\alpha}(\mathbf{k}) v_{i,\beta}(\mathbf{k}) \epsilon(\textbf{k}, T),
\end{aligned}
\end{equation}
where
\begin{equation}
    \epsilon(k, T) = \frac{\hbar \Omega_{i}(\mathbf{k})}{k_B T} \frac{e^{\hbar \Omega_{i}(\mathbf{k})/k_B T}}{(e^{\hbar \Omega_{i}(\mathbf{k})/k_B T} - 1)^2}.
\end{equation}

The integral is over the $n$-dimensional Brillouin zone (monolayer $n=2$, bulk $n=3$), and the sum $i \in (A, B)$ runs over all magnon–polaron modes. Here $T$ is the temperature in the sample, $\mathbf{k}$ is the wavevector, $\Omega_{i}(\mathbf{k})$ is the hybridized magnon-polaron mode energy, $v_{i,\beta}(\mathbf{k}) = \partial \Omega_{i}(\mathbf{k}) / \partial k_{\beta}$ is its group velocity, $\tau_{i}(\mathbf{k})$ the magnon–polaron lifetime, $\mathcal{S}_{i}(\mathbf{k}) = \pm 1$ is the spin polarization corresponding to magnon mode, and $W_{i}(k)$ is the magnonic weight, or the magnetic amplitude, of the $i^{th}$ magnon-polaron mode, given by the sum of the magnonic components of the corresponding eigenvector of the hybrid magnon–phonon Hamiltonian \cite{flebus2017magnon}.

We compute $S_{xx}$ and $S_{yy}$ by evaluating (4) using the dispersions from Eq. (2) and including magnon–phonon scattering via Fermi’s golden rule. Details of the calculation of $S_{\alpha \beta}$ can be found in the supplementary information; here we show the main results.

To obtain them, we adopt the parameter set $J_1 = 1~\mathrm{meV}$, $J_2 = -1~\mathrm{meV}$, $D = -0.1~\mathrm{meV}$, and $S = \tfrac{3}{2}$. The exchange and anisotropy constants are chosen to be representative, lying within the same order of magnitude as those reported for antiferromagnets \cite{dzian2025antiferromagnetic} and candidate altermagnetic materials \cite{vsmejkal2023chiral}. In the antiferromagnetic limit, $J_2' = J_2$, while for the altermagnetic case, we set $J_2' = -2$~meV. The phonon speed is set to $c = 3500$~m/s. First, we focus on the 2D limit. Fig. 2 shows the spin Seebeck coefficient in the magnetic sample, \( S_{\alpha \alpha}^{2D} \), plotted vs. the applied magnetic field $H$, in the 2D limit and the absence of phonons. For the antiferromagnetic configuration $(J'_{2} = J_{2})$, the longitudinal SSE is non-directional, i.e.,  \( S_{xx} = S_{yy} \) for all values of the external magnetic field (Fig. 2(a)). In contrast, in the altermagnetic configuration $(J'_{2} \neq J_{2})$, the longitudinal SSE is highly directional, i.e., the SSE is different for the same magnitude of magnetic field applied along $x$- and $y$-directions (Fig. 2 (b)). Moreover, we find that this directional asymmetry becomes more pronounced with increasing magnitude of the magnetic field. This enhancement arises directly from the anisotropy in the group velocities along the \( x \)- and \( y \)-directions in altermagnets. The directional asymmetry in the SSE with respect to the applied magnetic field is a signature of altermagnetic character.

\begin{figure}[]
 \begin{center}
    \begin{subfigure}{0.23\textwidth}
    \includegraphics[width=\linewidth]{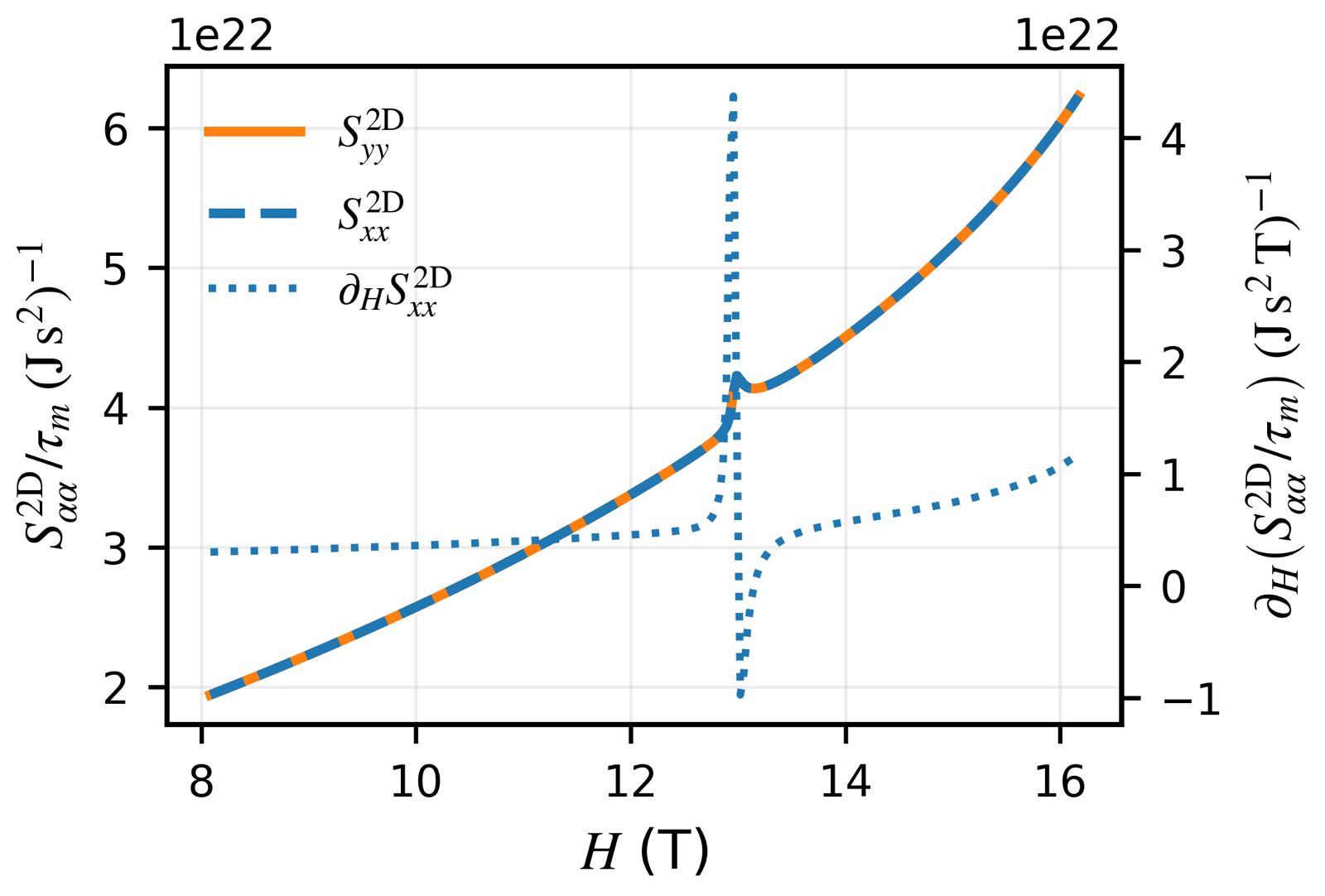}
    \caption{} 
  \end{subfigure}
  \begin{subfigure}{0.23\textwidth}
    \includegraphics[width=\linewidth]{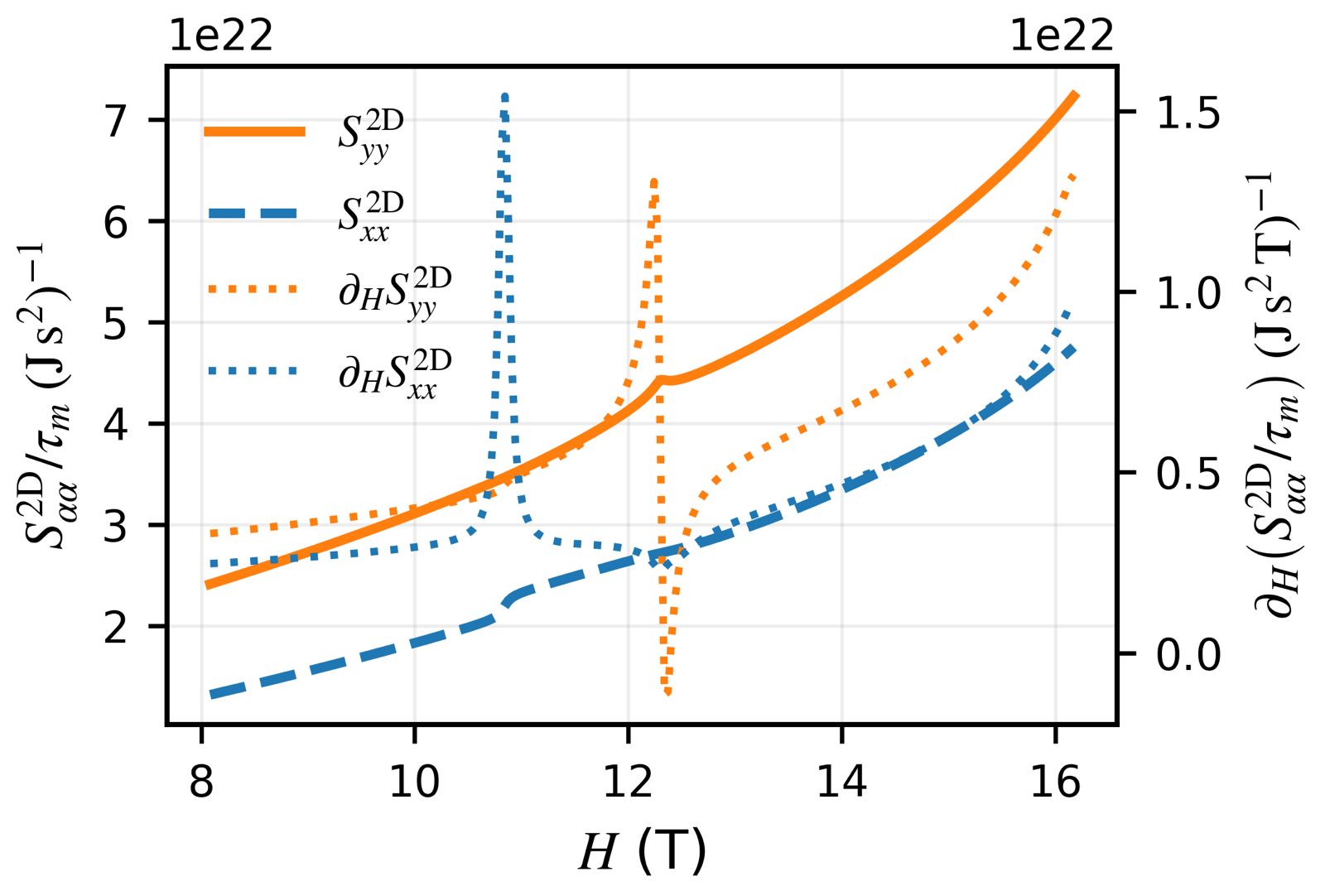}
    \caption{} 
  \end{subfigure}
  \begin{subfigure}{0.23\textwidth}
    \includegraphics[width=\linewidth]{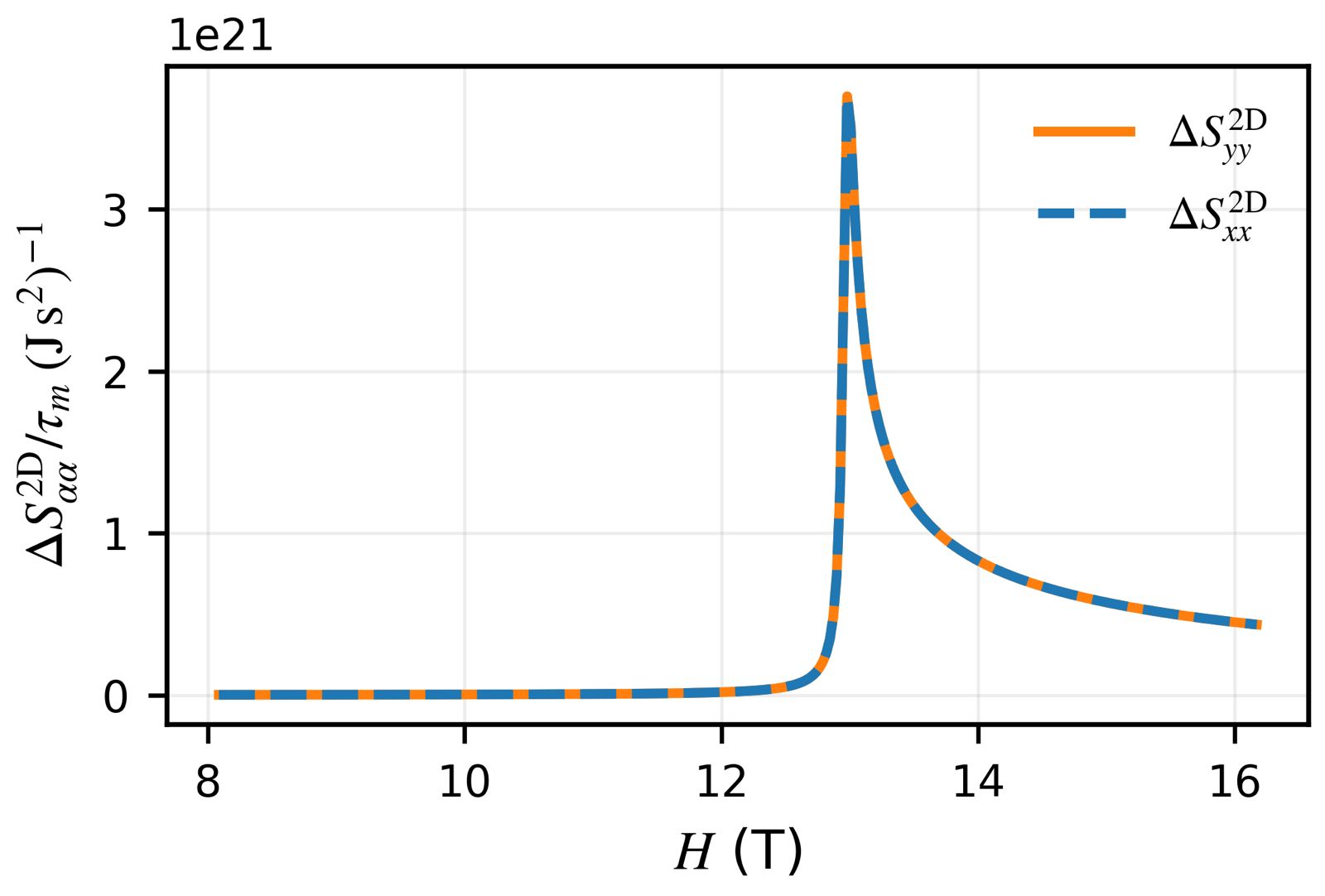}
    \caption{} 
  \end{subfigure}
  \begin{subfigure}{0.23\textwidth}
    \includegraphics[width=\linewidth]{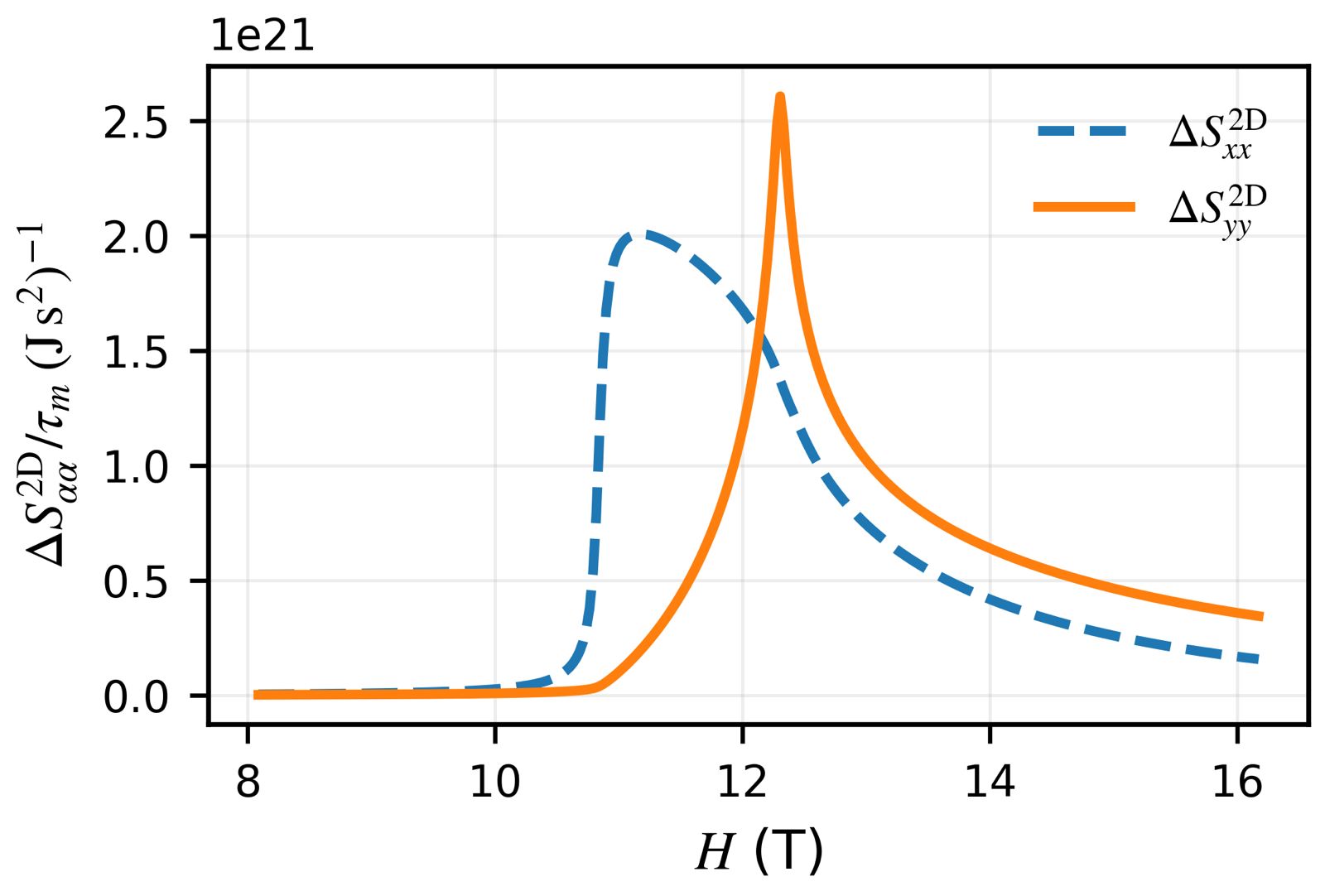}
    \caption{} 
  \end{subfigure}
  \end{center}
    \begin{center}
 \caption{\centerlast SSE vs. applied magnetic field in-plane along the $x$ and $y$ directions with magnetoelastic coupling for (a) an antiferromagnetic lattice and (b) for an altermagnetic lattice in the monolayer limit. Peaks in SSE are seen at the same magnetic field for the antiferromagnetic case in both $S^{2D}_{xx}$ and $S_{yy}^{2D}$. For the altermagnetic case, the peak in $S_{xx}^{2D}$ appears at a magnetic field different from the peak in $S_{yy}^{2D}$. The features are seen clearly by calculating the difference in SSE due to phonons (c) $\Delta S_{xx}$ and (d) $\Delta S_{yy}$ vs. magnetic field. $\tau_{m}$ is the magnon lifetime.}
    \end{center}
 \end{figure}
As discussed earlier, the inclusion of phonons can enhance the directional character of the system. We now proceed to calculate the SSE in both an antiferromagnet and an altermagnet in the presence of acoustic phonons. We do not consider the coupling to optical phonons, as their hybridization with magnons occurs at high energies and thus contributes negligibly to the SSE. The acoustic modes have a linear and symmetric dispersion with respect to the wavevector $\textbf{k}$. The details of the magnetoelastic coupling and acoustic phonons are shown in the supplementary information. Fig. 3 shows the SSE in a 2D magnet in the presence of phonons with respect to the magnetic field. Fig. 3 (a) shows the longitudinal SSE in the $x$ $(S_{xx})$ and $y$ $(S_{yy})$ directions, and the corresponding derivatives with respect to the applied magnetic field in the antiferromagnetic limit. We observe a pronounced peak in both $S_{xx}$ and $S_{yy}$. This peak arises at the point where the magnon frequency is equal to the phonon frequency, and the magnons and phonons hybridize to form magnon-polaron bands. The enhancement in the SSE due to magnon-polarons is well-known \cite{cornelissen2017nonlocal, kikkawa2016magnon,flebus2017magnon}. Since the mode crossing is symmetric with respect to the wavevector $k$, the peaks in $S_{xx}$ and $S_{yy}$ appear at the same magnitude of the magnetic field. However, in the altermagnetic limit (Fig. 3 (b)), the magnon dispersion is asymmetric with respect to $k$. As a result, the magnon bands and phonon bands cross at different magnetic fields along different wavevector. Thus, we observe the peaks at different values of magnetic field. This behavior is unique to altermagnets. Fig. 3 (c) and Fig. 3 (d) show the difference gained in the SSE due to the hybridization with phonons, $\Delta S^{2D}_{xx}$ and $\Delta S^{2D}_{yy}$. The positions of the peaks are seen much clearly in this case. The peak appears at $H \sim 11$T in $\Delta S_{xx}$ but at $H \sim 13$T in $\Delta S_{yy}$ for the altermagnetic case, reflecting the anisotropy in the SSE. In the antiferromagnetic limit, the peak appears at $H \sim 13$T for both $S_{xx}^{2D}$ and $S_{yy}^{2D}$. Exact point-touching is not required, as the magnetoelastic coupling hybridizes the modes over a finite momentum window set by the anticrossing gap \cite{white1965diagonalization, flebus2017magnon}. This near-resonant region contributes significantly to the spin Seebeck response. As a result, even in the altermagnetic case, where the resonance occurs only at isolated points, the magnon-polaron contribution remains sufficiently strong to generate clearly observable peaks. For the same reason, the magnon–polaron signal deviates from the bare magnon Seebeck response even at fields away from the exact touching condition.

\begin{figure}[]
 \begin{center}
    \begin{subfigure}{0.23\textwidth}
    \includegraphics[width=\linewidth]{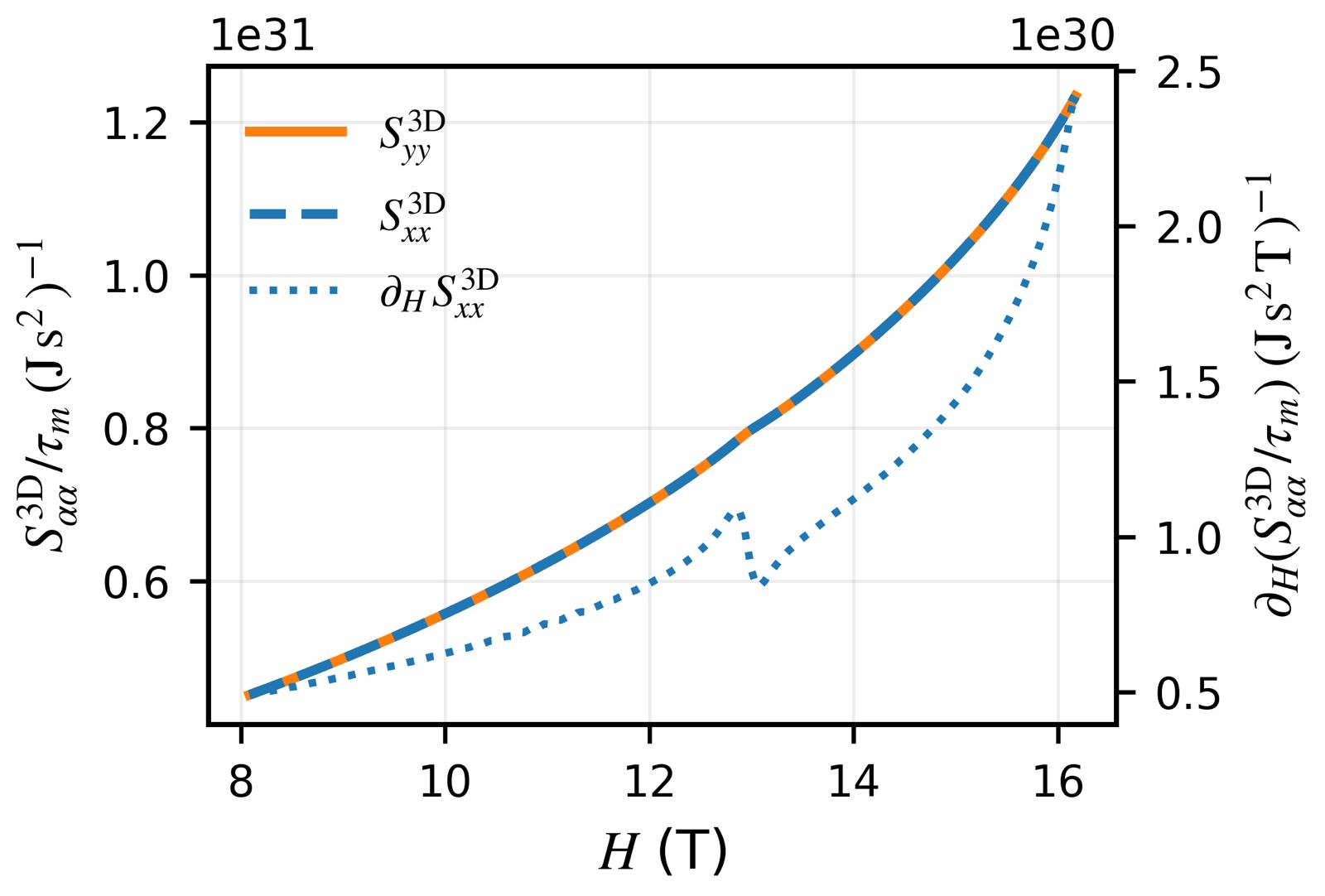}
    \caption{} 
  \end{subfigure}
  \begin{subfigure}{0.23\textwidth}
    \includegraphics[width=\linewidth]{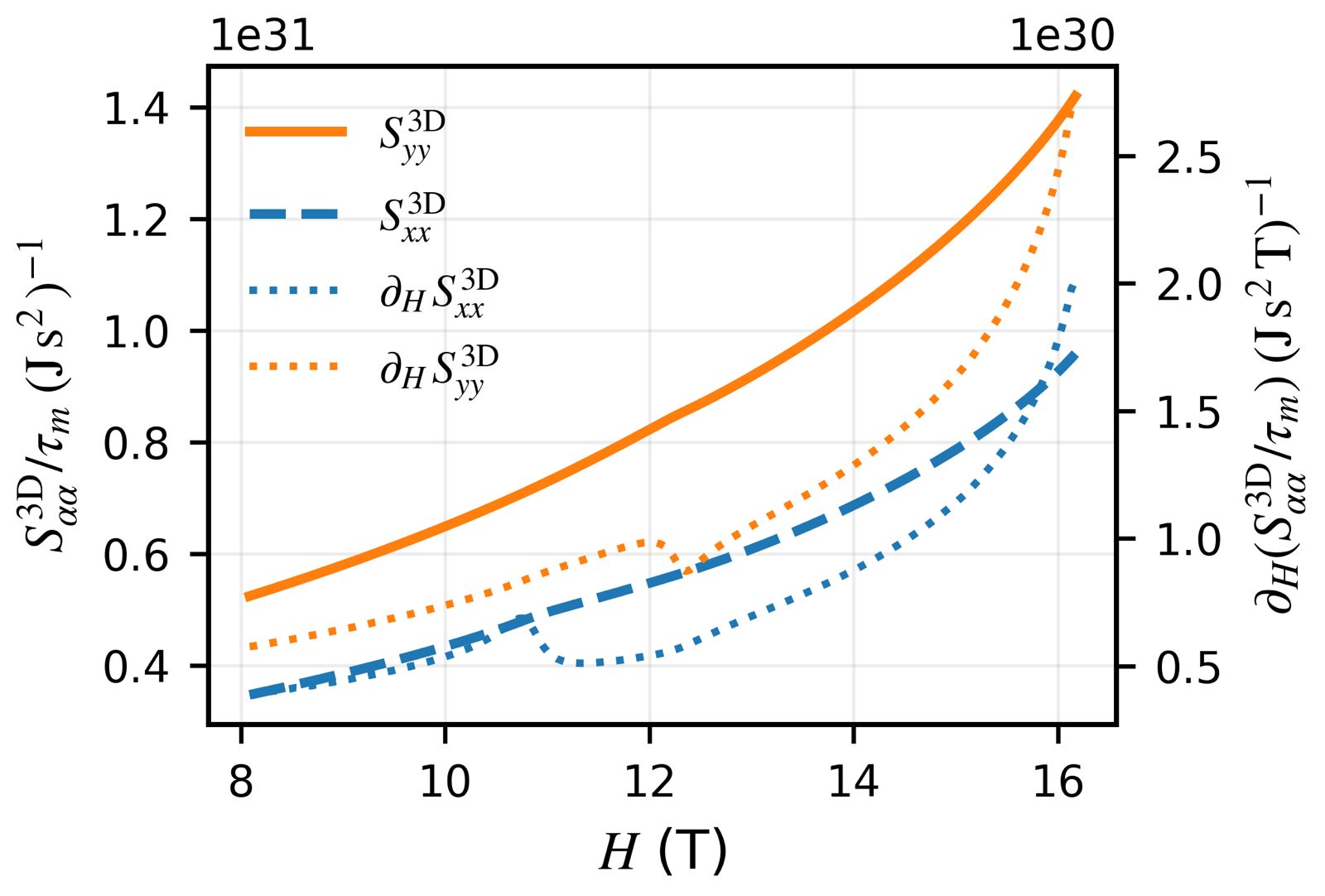}
    \caption{} 
  \end{subfigure}
  \end{center}
    \begin{center}
 \caption{\centerlast SSE vs. applied magnetic field along different $x$ and $y$ directions with magnetoelastic coupling for (a) an antiferromagnetic lattice and (b) for an altermagnetic lattice in the 3D bulk case.}
    \end{center}
 \end{figure}

Finally, we investigate a three-dimensional analog of our model, which is defined as layers of the 2D checkerboard stacked on top of each other with a weak interlayer exchange $J_{\text{inter}}$. In the bulk limit, we consider all phonon modes. We adopt realistic material parameters for the phonon velocities: \( c_{\lambda}^{x} = c_{\lambda}^{z} = 3500\,\text{m/s} \) for the two transverse modes, and \( c_{\lambda}^{y} = 7000\,\text{m/s} \) for the longitudinal mode. The parameters for the magnonic subsystem are kept the same as in the two-dimensional case, with the addition of a \( 0.01\,\text{meV} \) out-of-plane exchange coupling to account for the bulk geometry. The resulting SSE in the bulk configuration for the altermagnetic and antiferromagnetic limits is presented in Fig. 4.

The behavior in the bulk limit is similar to the behavior in the 2D limit. In Fig. 4 (a), we show the SSE $S_{xx}^{3D} = S_{yy}^{3D}$ and its derivative with respect to the magnetic field. The peaks are seen at the same strength of magnetic field along the $x$ and $y$ directions. Fig. 4 (b) shows the SSE along $x$ $(y)$ direction in the bulk limit given by $S_{xx}^{3D}$ $ (S_{yy}^{3D})$ for a bulk altermagnet. Similar to the 2D case, the peaks appear at different strengths of the magnetic field for $S_{xx}^{3D}$ and $S_{yy}^{3D}$. Thus, the asymmetry in the spin Seebeck response can be used to distinguish antiferromagnets from altermagnets even in the bulk limit.

In conclusion, we have demonstrated that the spin Seebeck effect provides a clear transport-based probe of altermagnetism. In contrast to antiferromagnets, where magnon degeneracy enforces isotropic SSE signatures, altermagnets exhibit a finite, strongly anisotropic response even at zero field. When phonons are included, the resulting magnon–polaron resonances further enhance this distinction. In antiferromagnets, a peak appears in the SSE at a fixed field along all directions, whereas in altermagnets the peak positions in the SSE along different directions appear at different field strengths, reflecting the underlying anisotropic magnon dispersion. These results establish that SSE measurements, particularly their directional dependence and magnon–polaron mediated features, can serve as robust hallmarks of altermagnetism. Beyond their fundamental significance in classifying spin-ordered states, our findings open the way to exploiting altermagnets as field-robust, stray-field-free spin caloritronic materials. Although our analysis focused on a d-wave altermagnet toy model, the mechanism is general and can be extended to other altermagnetic symmetries.

Our analysis highlights both constraints and opportunities for observing multiple SSE peaks in altermagnets. The effect requires for magnon and phonons to cross near resonance and for there to be splitting near resonance, conditions that can occur in known materials such as MnF$_2$ \cite{lovesey2026altermagnetism, morano2025absence}. We highlight the potential advantages of our method over neutron scattering in the supplementary information by focusing on MnF$_{2}$  (see section IV of the supplementary information).

Acknowledgements - R. D. and Y. M. B. acknowledge support from the project "Ronde Open Competitie ENW pakket 21-3" (file number OCENW.M.21.215) which is (partly) financed by the Dutch Research Council (NWO).

\bibliography{myreferences}

\end{document}


\title{Supplementary information for Magnon-polaron mediated spin Seebeck effect in altermagnets}

\author{Ilia Moghayer}
\affiliation{%
Kavli Institute of Nanoscience, Delft University of Technology, Lorentzweg 1, 2628 CJ, Delft, The
Netherlands
}%

\affiliation{Institute for Theoretical Physics, ETH Zurich, CH-8093 Zurich, Switzerland}%

\author{Ritesh Das}
\email{R.D.Das@tudelft.nl}
\affiliation{%
Kavli Institute of Nanoscience, Delft University of Technology, Lorentzweg 1, 2628 CJ, Delft, The
Netherlands
}%

\author{Yaroslav M. Blanter }
\email{Y.M.Blanter@tudelft.nl}
\affiliation{%
Kavli Institute of Nanoscience, Delft University of Technology, Lorentzweg 1, 2628 CJ, Delft, The
Netherlands
}%

\date{\today}

\maketitle

\section{magnon dispersion}

In this section, we analytically compute the magnon dispersion in our toy model using the spin-wave approximation. We consider small dynamic deviations of the sublattice magnetizations from their equilibrium orientations. These oscillations vary in space and time and are described by well-defined wavevectors \( \mathbf{k} \), leading to spin-wave excitations. We relate the spin operators \textbf{S} to the magnetic moment \( \vec{m}_{a,i}\), using the gyromagnetic ratio $\gamma$, as $\vec{m}_{a,i} = \frac{\gamma \hbar \mathbf{S}_{a,i}}{M_s}$, allowing us to rewrite Eq. 1 of the main text as \cite{Rezende2019, stancil2009spin}

\begin{align} 
\frac{E}{M_s} &= 
\hspace{0.1cm}\frac{S}{\gamma \hbar} J_1 \sum_{\langle  i,j \rangle} \vec{m}_{a,i} \cdot \vec{m}_{b,j} \notag - \sum_{i} (\vec{m}_{a,i} + \vec{m}_{b,i})\cdot H_0
\\ &+ \frac{S}{\gamma \hbar} \Big( J_2' \sum_{\langle \langle i,j \rangle \rangle_{y}} \vec{m}_{a,i} \cdot \vec{m}_{a,j}  \notag 
 + J_2'\sum_{\langle \langle i,j \rangle \rangle_{x}} \vec{m}_{b,i} \cdot \vec{m}_{b,j} \Big) \notag
\\ & + \frac{S}{\gamma \hbar} \Big( J_2 \sum_{\langle \langle i,j \rangle \rangle_{y}} \vec{m}_{b,i} \cdot \vec{m}_{b,j} +  J_2 \sum_{\langle \langle i,j \rangle \rangle_{x}} \vec{m}_{a,i} \cdot \vec{m}_{a,j}\Big)   \notag \\ 
\quad &+ \frac{S}{\gamma \hbar} D \sum_{i}[ (m_{b,i}^{y})^2 +(m_{a,i}^{y})^2 ]
.
\label{eq:energy_density}
\end{align}

For a spin aligned along the $y$-axis in its ground state (see Fig. 1 in the main text), the dynamic magnetic moment of its excited state is written in the spin-wave approximation as:

\begin{equation}
\vec{m}_{a,i} = \mathbf{\hat{y}} m_{a,i}^{y} + \left( \mathbf{\hat{x}} m_{a,i}^{x} + \mathbf{\hat{z}} m_{a,i}^{z} \right) e^{-i \omega t}  e^{i \mathbf{k} \cdot \mathbf{r}_{i}},
\label{eq:generalMagnetization}
\end{equation}
where $\vec{m}_{a, i}$ is the magnetization vector at site $i$ in sublattice $a$, $m^{n}_{a,i}$ is the component of the magnetization along $n$ axis at site $i$ in sublattice $a$, $S$ is the spin, and $\omega$ is the frequency. Furthermore, $\textbf{k}$ is the wavevector and $\textbf{r}_{i}$ is the spatial vector of lattice site $i$. We substitute the dynamic magnetization from Eq. (2) into the energy expression given in Eq. (1). We proceed by computing the effective field and filling it in into the Landau-Lifshitz equation \cite{Rezende2019,Rezende2020}, which yields the eigenvalue matrix

\begin{equation}
\mathbf{M} = \gamma
\begin{pmatrix}
0 & 0 & L_1 + X_1 & X \\
0 & 0 & -X & L_2 - X_2 \\
-L_1 - X_1 & -X & 0 & 0 \\
X & -L_2 + X_2 & 0 & 0
\end{pmatrix},
\end{equation}

with: 

\begin{subequations}
\begin{align}
L_1 &= H_0 + H_{E1} + H_A + 2H_{E2} + 2H_{E2'}, \\ 
L_2 &= H_0 - H_{E1} - H_A - 2H_{E2} - 2H_{E2'},\\ 
X &= H_{E1} \cdot \gamma_3, \\
X_1 &= 2 (H_{E2} \cdot \gamma_1 + H_{E2'} \cdot \gamma_2), \\
X_2 &= 2(H_{E2} \cdot \gamma_2 + H_{E2'} \cdot \gamma_1).
\end{align}
\label{eq:matrix_definition}
\end{subequations}
The \( \gamma_i \) factors correspond to the structure factors taken from \cite{Cui2023}, and 

\begin{subequations}
\begin{align}
    H_{E1} &= \frac{2S z_1 J_1}{\gamma \hbar}, \\ H_{E2} &= \frac{2S z_2 J_2}{\gamma \hbar}, \\ H_{E2'} &= \frac{2S z_2 J_{2}'}{\gamma \hbar}, \\ H_A &= \frac{2S D}{\gamma \hbar}.
    \label{eq:fields}
\end{align}
\end{subequations}
Here, \( z_i \) represents the coordination number, which specifies the number of nearest-neighbor interactions per spin site. In this model, \( z_1 = 4 \) and \( z_2 = 2\) denote the coordination numbers for inter-sublattice and intra-sublattice interaction, respectively. The altermagnetic behavior is visible in Eq. (4) from the difference between $X_1$ and $X_2$. For an antiferromagnet, $X_{1} = X_{2}$.

\begin{figure}
    \centering
    \begin{minipage}{0.23\textwidth}
        \centering
        \includegraphics[width=\linewidth]{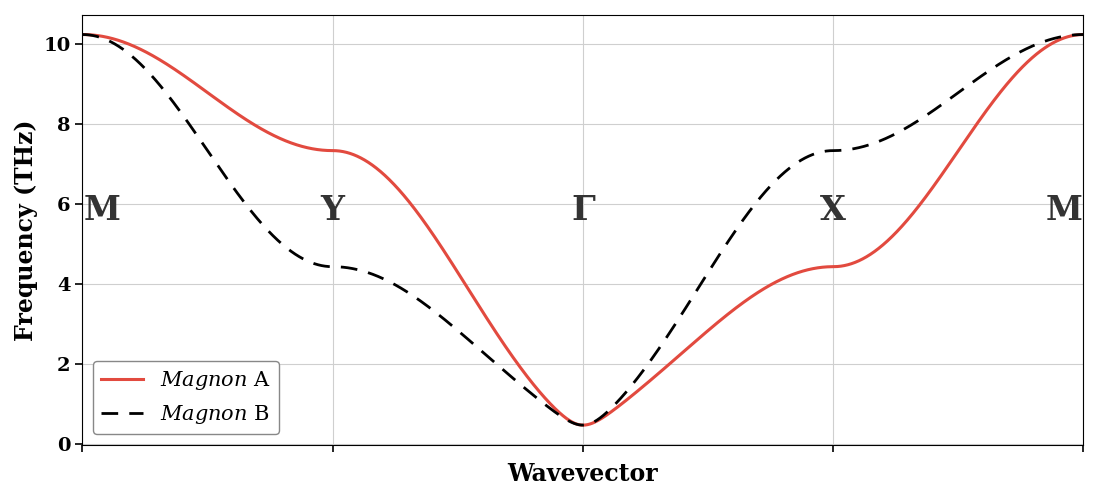}
    \end{minipage}
    \hfill
    \begin{minipage}{0.23\textwidth}
        \centering
        \includegraphics[width=\linewidth]{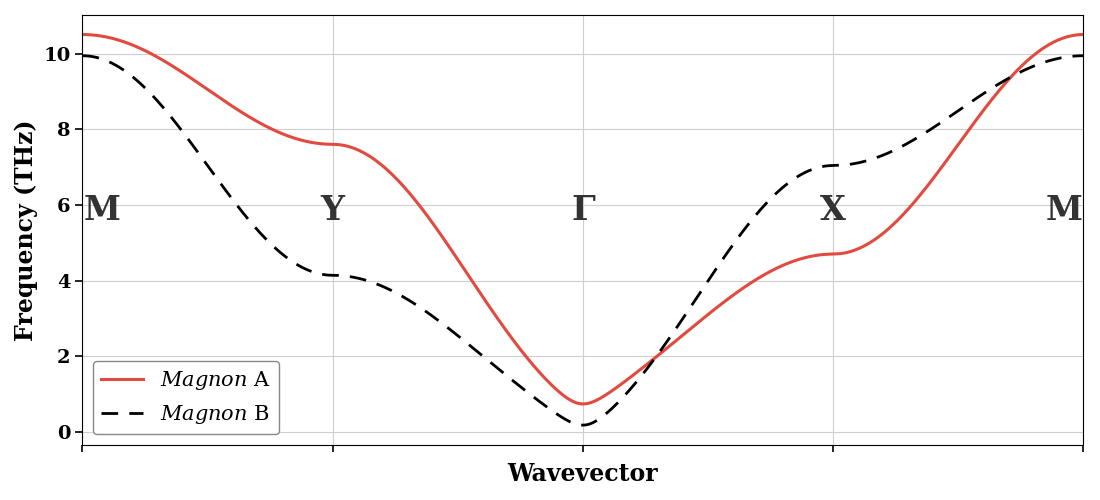}
    \end{minipage}
    
    \begin{minipage}{0.23\textwidth}
        \centering
        \includegraphics[width=\linewidth]{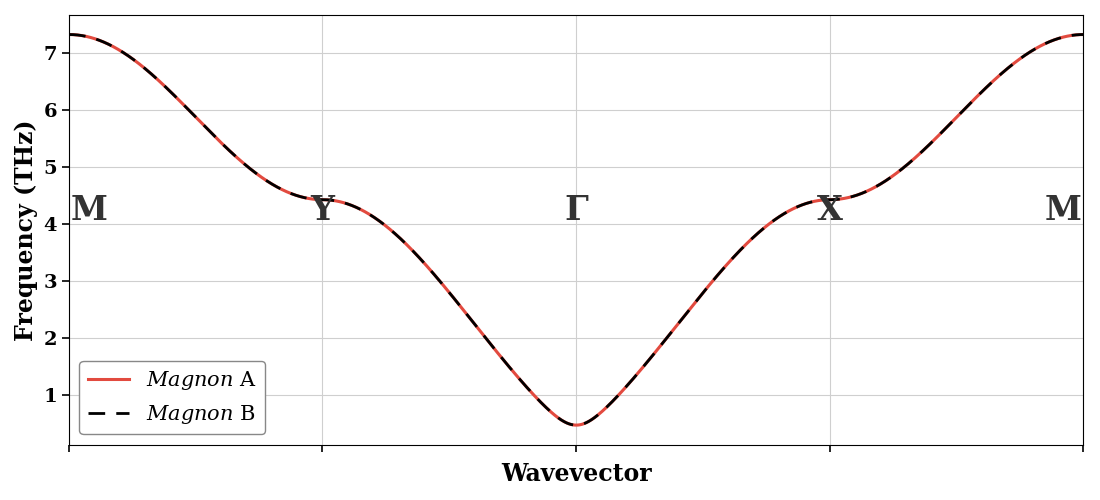}
    \end{minipage}
    \hfill
    \begin{minipage}{0.23\textwidth}
        \centering
        \includegraphics[width=\linewidth]{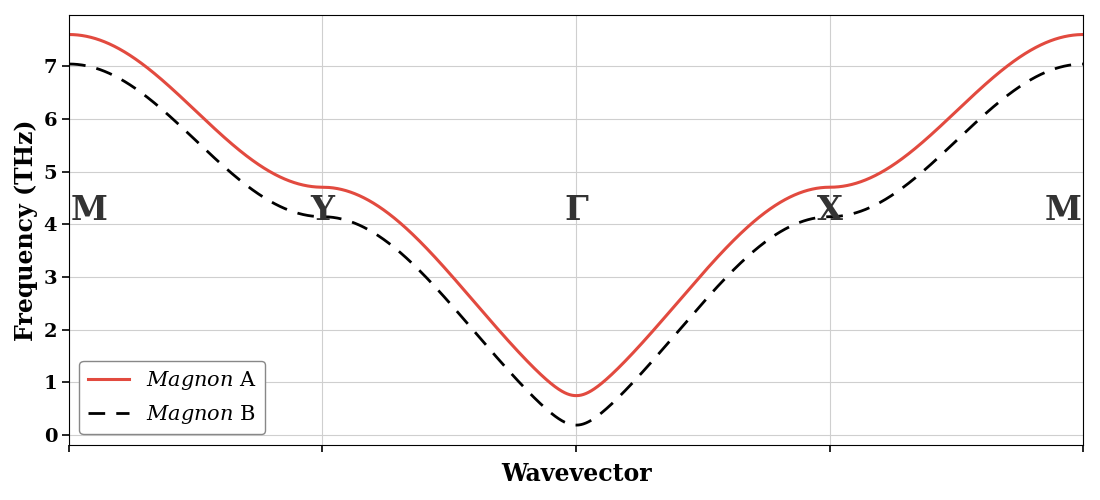}
    \end{minipage}

    \caption{Magnon dispersion of the toy model used for antiferromagnetic and altermagnetic condition with and without an external field. 
The top row displays the magnon dispersion relations in the altermagnetic condition \( J_2 \neq J_2' \), where the left panel corresponds to zero external magnetic field, and the right panel shows the dispersion under an applied field. 
The bottom row presents the dispersion under the assumption that \( J_2 = J_2' \), effectively recovering an antiferromagnetic state. 
In the absence of a field (bottom left), the magnon bands remain degenerate, whereas applying an external field (bottom right) results in a uniform Zeeman splitting, distinct from the behavior observed in the altermagnetic phase. 
The dispersion is plotted along the high-symmetry path \( \text{M} \left(\frac{\pi}{a}, \frac{\pi}{b}\right) \to \text{Y} \left(0, \frac{\pi}{b}\right) \to \Gamma (0,0) \to \text{X} \left(\frac{\pi}{a}, 0\right) \to \text{M} \left(\frac{\pi}{a}, \frac{\pi}{b}\right) \).}
    \label{fig:AFM_combined_all}
\end{figure}

We now solve the eigenvalue problem, which ultimately yields the following magnon dispersion relation:

\begin{equation}
\omega_{A,B} = \frac{ \mp (\Lambda_2 - \Lambda_1) + \sqrt{ (\Lambda_2 - \Lambda_1)^2 + 4(\Lambda_1 \Lambda_2 - X^2)} }{2}
\end{equation}
where:

\begin{equation}
\begin{aligned}
\Lambda_1 &= H_0 + H_{A} + H_{E1} - 2(H_{E2} + H_{E2'})\\ &+ 2(H_{E2} \gamma_1 + H_{E2'} \gamma_2), \\
\Lambda_2 &= -H_0 + H_{A} + H_{E1} - 2(H_{E2} + H_{E2'}) \\&+ 2(H_{E2} \gamma_2 + H_{E2'} \gamma_1).
\end{aligned}
\end{equation}
The magnon dispersion is presented in Fig.~\ref{fig:AFM_combined_all} for \( H = 0 \, \mathrm{T} \) and \( H = 10 \, \mathrm{T} \). The top row shows the dispersion in the altermagnetic case ($J_{2} \neq J'_{2}$). The bottom row shows the antiferromagnetic case ($J_{2} = J'_{2}$). We focus on the altermagnetic dispersion first. In the absence of an external magnetic field, the two magnon bands remain split: a feature that distinguishes this system from a conventional antiferromagnets. These excitations are known as chiral magnons \cite{vsmejkal2023chiral}, characterized by a directional asymmetry in their magnonic dispersion relation.

Upon applying an external magnetic field, we observe a behavior analogous to that in antiferromagnets: the energy of the \textit{A} mode increases, while the energy of the \textit{B} mode decreases. This results in the emergence of nodal points near \( \Gamma \). A similar effect has been previously reported by Alaei \textit{et al.} \cite{alaei2025origin}. For \( J_2 = J_2' \), we recover a conventional antiferromagnetic state with characteristic antiferromagnetic magnon dispersion (see Fig. 1 bottom row). As expected, in this scenario, the alternating band structure observed in the altermagnetic case disappears, and the two magnon bands become degenerate in the absence of an external field. Additionally, under a finite external field, the system exhibits an isotropic Zeeman splitting, in contrast to the field-induced anisotropic modifications in the altermagnetic case.

\begin{figure}
    \centering
    \begin{minipage}{0.23\textwidth}
        \centering
        \includegraphics[width=\linewidth]{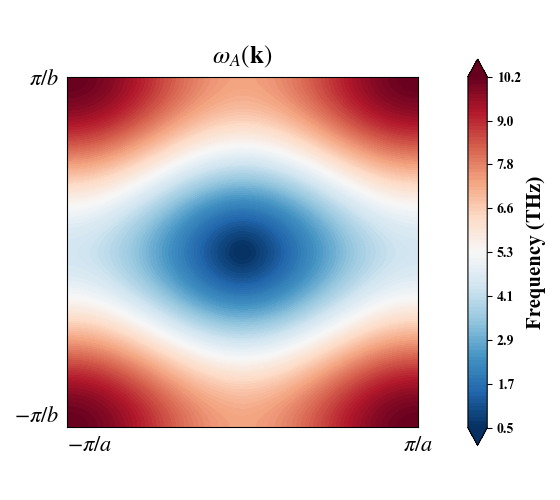}
    \end{minipage}
    \hfill
    \begin{minipage}{0.23\textwidth}
        \centering
        \includegraphics[width=\linewidth]{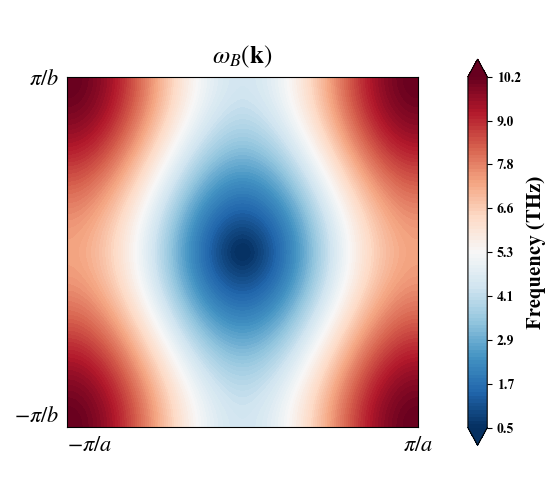}
    \end{minipage}

    \caption{Contour plot of the magnon dispersion in the toy model in the altermagnetic condition. The left panel represents the dispersion of the \( A \) mode, while the right panel shows the \( B \) mode.}
    \label{fig:contour}
\end{figure}

The contour plots of both magnon bands at zero external field are displayed in Figure \ref{fig:contour}. Unlike in antiferromagnets, where the contour is typically isotropic, we observe an anisotropic structure due to the presence of the symmetry-breaking condition \( J_2 \neq J_2' \). 

\subsection{Three-Dimensional Case }

To extend our toy model to 3D, a ferromagnetic exchange coupling out of plane is added to the monolayer Hamiltonian:

\begin{equation}
    \hat{H}_{\text{bulk}} = \hat{H}_{\text{mono}} + J_{\text{inter}} \sum_{\langle i,j \rangle_{\perp}} (\mathbf{S}_{a,i} \cdot \mathbf{S}_{a,j} + \mathbf{S}_{b,i} \cdot \mathbf{S}_{b,j}),
\end{equation}

\noindent
where the sum runs over nearest neighbors along the out-of-plane direction, and $J_z$ denotes the interlayer exchange constant. This term introduces coupling between adjacent layers with lattice constant $c$ and modifies the magnon dispersion by incorporating the $z$-component of the crystal momentum $\textbf{k}_{z}$. The extended structure factor is defined as:
\begin{align}
    \gamma_{4} &= \cos(\textbf{k}_z c).
\end{align}

\noindent
The effective third-neighbor exchange field becomes:
\begin{equation}
    H_{E3} = \frac{S z_3 J_{\text{inter}}}{\gamma \hbar}.
\end{equation}

The magnon dispersion relations are then derived very similarly to the monolayer case, and are given by:
\begin{align}
    \omega_A(\mathbf{k}) &= \frac{1}{2} \left[ -(Z_2 - Z_1) + \sqrt{(Z_2 - Z_1)^2 + 4(Z_1 Z_2 - X^2)} \right], \\
    \omega_B(\mathbf{k})  &= \frac{1}{2} \left[  (Z_2 - Z_1) + \sqrt{(Z_2 - Z_1)^2 + 4(Z_1 Z_2 - X^2)} \right],
\end{align}

\noindent
with:
\begin{subequations}
\begin{align}
    Z_1 &= Y_1 + X_1, \\
    Z_2 &= - Y_2 + X_2, \\
    X   &= \gamma H_{\text{inter}} \gamma_{3}, \\
    Y_1 &= \gamma \left( H_{\text{inter}} - 2 H_{\text{intra}} + H_A + H_0 \right), \\
    Y_2 &= \gamma \left( -H_{\text{inter}} + 2 H_{\text{intra}} - H_A + H_0 \right), \\
    X_1 &= 2\gamma \left( H_{E2} \gamma_{1} + H_{E2}' \gamma_{2} + H_{E3} \gamma_{4} \right), \\
    X_2 &= 2\gamma \left( H_{E2} \gamma_{2} + H_{E2}' \gamma_{1} + H_{E3} \gamma_{4} \right).
\end{align}
\end{subequations}

\section{linear phonon bands in 2D}
For a monolayer, the out-of-plane (ZA) phonon band has a quadratic dispersion relation, \( \omega(\textbf{k}) \propto \textbf{k}^2 \). However, in van der Waals (vdW) magnets under tension, this mode can harden due to additional energy costs. This can be captured by the total strain energy, which receives contributions from several terms. First, the pure bending energy is given by
\begin{equation}
    E_0 = \frac{\kappa}{2} \int \mathrm{d}^2\mathbf{r} \left[ \nabla^2 h(\mathbf{r}) \right]^2,
\end{equation}
where \( h(\mathbf{r}) \) is the vertical displacement field, and \( \kappa \) is the bending stiffness. This term dominates in the case of smooth distortions and reflects the energy cost of curvature.
Second, an effective tension term contributes,
\begin{equation}
    E_T = \frac{\Gamma}{2} \int \mathrm{d}^2\mathbf{r} \left[ \nabla h(\mathbf{r}) \right]^2,
\end{equation}
where \( \Gamma \) represents a tension coefficient, which may originate from van der Waals interlayer coupling or mechanical coupling to a substrate.

Transforming to momentum space, the resulting dispersion for the flexural phonon mode, as given by Neto \textit{et al.}
\cite{castro2009electronic}:

\begin{equation}
    \omega(\textbf{k}) = \textbf{k} \sqrt{ \frac{\Gamma}{\rho} + \frac{\kappa}{\rho} \textbf{k}^2 },
\end{equation}
where \( \rho \) is the area mass density of the membrane. In the limit of strong tension \( \Gamma \gg \kappa \textbf{k}^2 \), the dispersion becomes linear \( \omega(\textbf{k}) \sim \textbf{k} \), indicating significant linearization of the acoustic out-of-plane mode, consistent with the findings of  Iakovlev \textit{et al.}\cite{iakovlev2022flexural}. Here, we consider the limit where either interlayer van der Waals forces are strong enough or there is strong external tension to linearize the ZA mode at long wavelengths.

\section{Magnon-Polaron Mediated Spin
transport}

A magnon spin current model has been developed for the spin Seebeck effect in altermagnets~\cite{Cui2023}, and for magnon-polaron-mediated spin transport in YIG~\cite{flebus2017magnon}. We extend these models to the case of antiferromagnetic and altermagnetic magnon-polaron mediated spin transport phenomena.
As a framework to study the relevant transport mechanisms, we employ the Boltzmann equation. For this semiclassical approach to be valid, the magnon-polaron modes must be well-defined quasiparticles. This condition can be expressed as:
\begin{equation}
\tau^{-1}_{i}(\textbf{k}) \ll \Delta \omega_{i}(\textbf{k}),
\end{equation}
where \( \Delta \omega_{i}(\textbf{\textbf{k}}) \) denotes the energy gap between magnon-polaron branches, and $\tau$ is the scattering time of the magnon-polaron in mode $i$. When this condition holds, the hybrid modes are well separated and maintain their coherence.

Assuming this, the number of magnon-polarons in the \( i^{\text{th}} \) mode within the magnetic volume is given by the Bose-Einstein distribution:
\begin{equation}
n^{0}_{i}(\textbf{k}) = \frac{1}{e^{\frac{\hbar \Omega_{i}(\textbf{k})}{k_B T}} - 1}.
\end{equation}
Where $\Omega_{i}$ is the frequency of the hybridized magnon-polaron mode. Here we assume that the magnon distribution thermalizes efficiently to the phonon temperature, resulting in a locally equilibrated system with both magnons and phonons at temperature $T$.

For this assumption to hold, magnons must equilibrate their energy much faster than they decay via spin-non-conserving processes. This leads to the condition:
\begin{equation}
\tau_{\text{c}}^{-1} \gg \tau_{\text{sp}}^{-1},
\end{equation}
where \( \tau_{\text{c}}^{-1} \) is the spin-conserving relaxation rate, and \( \tau_{\text{sp}}^{-1} \sim g \omega \) is the spin-non-conserving relaxation rate, with \( g \) denoting the Gilbert damping parameter. This separation of timescales is one of the main reasons YIG is widely used in spin Seebeck experiments, as its Gilbert damping is relatively low \cite{flebus2017magnon}.

From Rezende \textit{et al.}~\cite{rezende2016theory}, we have that \( \vec{\mathcal{J}}_s^z \), the flux of spin angular momentum through the surface for antiferromagnetic order, is given by:

\begin{equation}
\vec{\mathcal{J}}_s^z = -\frac{\hbar}{V} \sum_\textbf{k} \vec{v}_{m}(\textbf{k}) \left( \hat{a}_\textbf{k}^{\dagger} \hat{a}_\textbf{k} - \hat{b}_\textbf{k}^{\dagger} \hat{b}_\textbf{k} \right),
\end{equation}
where \( \vec{v}_{mk} \) is the group velocity of the magnon, $V$ is the volume, and $\hat{a}^{(\dagger)}$ and $\hat{b}^{(\dagger)}$ represent the annihilation (creation) operators of the magnon-polaron modes. The phonons carry linear momentum but not spin angular momentum, since we are interested in the spin transport, we must isolate the magnon contribution of the hybrid magnon-phonon excitations.

To account for this, we introduce a magnonic weight \( W_{i}(\mathbf{k}) \), which quantifies the magnonic character of each hybrid eigenmode \( i \) at momentum \( \mathbf{k} \). The explicit expression for \( W_{i}(\mathbf{k}) \) will be given after diagonalizing the full magnon-phonon Hamiltonian. In the continuum limit, we get

\begin{equation}
\vec{\mathcal{J}}_s^z = \frac{\hbar}{(2\pi)^n} \int d^n \textbf{k} \sum_i \vec{v}_{m}(\textbf{k}) W_{i,\textbf{k}} \mathcal{S}_{i}(\textbf{k}) \delta n_{i}(\textbf{\textbf{k}}),
\end{equation}
where  $\delta n_{i}(\textbf{k})(\vec{r}) = n_{i}(\textbf{k})(\vec{r}) - n^{0}_{i}(\textbf{k})$ the number in excess equilibrium, $\mathcal{S}_{i,\textbf{k}} = \pm 1$, the spin sign of mode $i$ at wave vector $\textbf{k}$, and $n$ is the dimensionality of the insulator.

The distribution of the magnon number under the influence of a thermal gradient can be calculated using the Boltzmann transport equation. In the absence of external forces, within the relaxation time approximation, and under steady-state conditions, the Boltzmann equation for small deviations from equilibrium takes the form:

\begin{equation}
\delta n_{i}(\textbf{k}) = -\tau_{i}(\textbf{k}) \vec{v}_{i}(\textbf{k}) \cdot \nabla n_{i}(\textbf{k})^{0} - \tau_{i}(\textbf{k}) \vec{v}_{i}(\textbf{k}) \cdot \nabla \left[\delta n_{i}(\textbf{k}) \right],
\label{boltzmann}
\end{equation}

The first term in Equation (22) represents the driving term, which in our case corresponds to the direct response to a temperature gradient. The second term accounts for the spin current arising from the spatial variation of the non-equilibrium magnon distribution. For the purposes of this study, we focus on the direct response to the temperature gradient.

We can write:
\begin{equation}
\nabla n_{i}^{0}(\textbf{k})^{} = \frac{\partial n_{i}^{0}(\textbf{k})^{}}{\partial T} \nabla T \quad.
\label{partial}
\end{equation}

As Eq. (22) is restricted for small deviations from equilibrium, the temperature should not vary significantly over the typical distance a magnon travels between scattering. We quantify this as: $
\frac{\nabla T}{T} \;\ll\; \frac{1}{\ell_{\mathrm{mfp}}}
$, with $\ell_{\mathrm{mfp}}$ being the mean free path of the magnon, which is typically on the order of 1 micron.

Filling in Eq. (23) into the Boltzmann equation, taking only into account the convective part yields: 
\begin{equation}
- \tau_{i}(\textbf{k}) \vec{v}_{i}(\textbf{k}) \cdot \nabla n_{i}^{0}(\textbf{k}) = -\tau_{i}(\textbf{k}) \vec{v}_{i}(\textbf{k}) \frac{\partial n^{0}_{i}(\textbf{k})} {\partial T} \cdot\nabla T.
\end{equation}

We now define the spin current density:
\begin{equation}
\vec{J}_{s \nabla T}^{\beta \hat{\alpha}} = S_{\alpha \beta} \frac{\partial T}{\partial \beta},
\end{equation}

\noindent
where \( \vec{J}_{s \nabla T}^{\beta \, {\alpha}} \) denotes the spin current density contribution along the \( {\beta} \)-direction due to a thermal gradient $\nabla T$ in the \( {\alpha} \)-direction, and \( S_{\alpha \beta} \) is the corresponding response coefficient for a temperature gradient in the \( \beta \)-direction. When \( {\alpha} = {\beta} \), the response is referred to as the longitudinal spin Seebeck effect (LSSE), whereas for \( {\alpha} \neq {\beta} \), it is known as the spin Nernst effect. This is expressed as:

\begin{equation}
\begin{aligned}
S_{\alpha\beta} = -\frac{\hbar}{(2\pi)^n k_B T} \int d^n \mathbf{k} \sum_{i} W_{i}(\mathbf{k}) \mathcal{S}_{i}(\mathbf{k}) \tau_{i}(\mathbf{k})\\
\times v_{i,\alpha}(\mathbf{k}) v_{i,\beta}(\mathbf{k}) \epsilon(\textbf{k}, T),
\end{aligned}
\end{equation}
where
\begin{equation}
    \epsilon(k, T) = \frac{\hbar \Omega_{i}(\mathbf{k})}{k_B T} \frac{e^{\hbar \Omega_{i}(\mathbf{k})/k_B T}}{(e^{\hbar \Omega_{i}(\mathbf{k})/k_B T} - 1)^2}.
\end{equation}

The scattering time \( \tau_i(\mathbf{k}) \) is given by

\begin{equation}
\tau_i(\mathbf{k}) =
\frac{\tau_{\text{m}} \, \eta \frac{v_g^{\text{m}}}{v_g^{\text{ph}}}}
{1 -(\eta \frac{v_g^{\text{m}}}{v_g^{\text{ph}}} - 1) |W_i(\mathbf{k})|}.
\end{equation}

Here, $\tau_{\text{m}}$ is the scattering time of the magnons, $v_{g}^{\text{m(ph)}}$ is the magnon (phonon) group velocity and $\eta$ is given by

\begin{equation}
    \eta = \, \frac{v_g^{\text{m}}}{v_g^{\text{ph}}} \times \frac{\tau^{\text{ph}}(\mathbf{k})}{\tau^{\text{m}}(\mathbf{k})} .
\end{equation}

\section{\ce{MnF$_{2}$}}
We compare the effectiveness of our method for a known altermagnetic candidate. MnF$_2$ provides a particularly compelling case study. Altermagnetism in MnF$_{2}$ is heavily debated \cite{lovesey2026altermagnetism, morano2025absence, faure2025altermagnetism}. Along the $\Gamma$--M and M--$\Gamma$ symmetry directions, the magnon dispersion is expected to exhibit altermagnetic spin splitting already at zero magnetic field. The underlying exchange topology is predominantly antiferromagnetic, supplemented by an effective ferromagnetic interlayer coupling and additional diagonal exchange interactions, rendering the system qualitatively similar to the three-dimensional toy model discussed above.

Importantly, previous neutron scattering experiments were unable to directly resolve this spin splitting. Instead, they established that the exchange parameters governing the altermagnetic behavior differ by less than the experimental resolution, corresponding to an upper bound of approximately $0.02$~meV. This limitation provides a natural benchmark for comparing the resolving power of neutron spectroscopy with that of phonon-mediated spin Seebeck effect (SSE) measurements in altermagnets.

To this end, we employ the numerical spin model of Morano \textit{et al.}~\cite{morano2025absence}, augmented by an external magnetic field applied along the out-of-plane direction. We set the exchange asymmetry to the experimental bound of $0.02$~meV. Using an estimate for the transverse acoustic phonon velocity near the magnon--phonon resonance of $v_{\mathrm{TA1}} \approx 1500$~m/s, we find that the difference in magnetic field required to satisfy the point-touching condition for the two magnon branches is on the order of $0.6$~T. Including a second transverse acoustic branch with velocity $v_{\mathrm{TA2}} \approx 2500$~m/s yields field separations on the order of $1$~T.

These field separations provide a direct route to estimating the exchange asymmetry via magnon--phonon resonance measurements. Our results therefore suggest that phonon-mediated SSE measurements can achieve a sensitivity comparable to neutron scattering, and may even become competitive, positioning this approach as a promising candidate for resolving the predicted spin splitting in MnF$_2$, contingent on achievable experimental field resolution.

That said, caution is warranted in generalizing this conclusion. The precise location of the point-touching conditions depends sensitively on both the phonon velocities and the detailed magnon dispersion, for which only approximate estimates are employed here. Moreover, even if MnF$_2$ proves favorable, other altermagnetic materials may not exhibit resolvable field splittings. The observability of magnon--phonon touching points depends critically on magnetic anisotropy: a sufficiently large anisotropy is required to open a gap and increase the spin-flop field, thereby producing a strongly curved magnon dispersion near resonance. This requirement applies equally to antiferromagnets and altermagnets, whereas ferromagnets naturally exhibit such curvature.

Notably, many candidate altermagnets discussed in the literature emphasize exchange-driven symmetry breaking while neglecting magnetic anisotropy. In the limit of weak uniaxial anisotropy, the altermagnetic phase may become unstable in finite magnetic field. While such approximations are justified when focusing on exchange-induced spin splitting, they highlight the importance of incorporating magnetic anisotropy in future searches for experimentally viable materials, particularly in the context of magnon--phonon-based probes under finite magnetic fields.

\bibliography{myreferences}